\documentstyle[12pt]{article}
\begin{document}
\title{{\large Sov. Phys. Usp. {\bf 34} (8), p. 717-728 August 1991}\\
High-temperature superconductivity and the characteristics of the
electronic energy spectrum}
\author{V. A. Moskalenko, M. E. Palistrant, and V. M. Vakalyuk\\
Institute of Applied Physics, Academy of Sciences of the Republic\\
of Moldavia}
\maketitle
\begin{abstract}
The possibility is indicated of applying the theory of superconductors with
overlap-\\ ping energy bands to describe the thermodynamic and electromagnetic
properties of the high-temperature compounds $La_{2-x}(Ba,Sr)_xCuO_4$ and
$YBa_2Ci_3O_{7-\delta}$. The two-band model was used to obtain high values of
$T_c$, two energy gaps $2\Delta_1/T_c > 3.5$ and $2\Delta_2/T_c < 3.5$, large
negative values of $d\ln T_c/d\ln V$ ($V$ is the volume) in lanthanum ceramics,
small values of the jump in the electron heat capacity at $T=T_c$, negative
curvature of the upper critical magnetic field $H_{c2}$ near the transition
temperature, etc. Such behavior of the above quantities is observed
experimentally. A description is also obtained of the decrease in $T_c$ as the
disordering of oxygen increases, and also as copper atoms are replaced by a
nonmagnetic impurity (Al, Zn, etc.). The main mechanism responsible for this
decrease is the interband scattering of electrons by impurities and by randomly
distributed oxygen vacancies. A theory has been developed of multiband
superconductors which takes into account the points of high symmetry in
momentum space. On the basis of this theory one can explain the existence of a
plateau in the dependence of $T_c$ on $\delta$ for $YBa_2Cu_3O_{7-\delta}$, and
also in the dependence of $T_c$ on $x$ for $La_{2-x}(Ba,Sr)_xCuO_4$, that has
been observed in a number of experiments. Moreover, this theory also explains
the presence of two maxima in the dependence of $T_c$ on pressure for
$Bi_2Sr_2CaCu_2O_8$.
\end{abstract}

\section{INTRODUCTION}
The discovery of high-temperature superconductivity in
$$
La_{2-x}(Ba, Sr)_{x}CuO_{4-\delta}(T_{c} \sim 40\ K),\
YBa_{2}Cu_{3}O_{7-\delta}(T_{c} \sim 90\ K)
$$
and other ceramics has stimulated experimental research on the properties of
these materials. In particular, the oxygen composition, the effect of
substitution of copper atoms by other metals, and the effect of neutron
irradiation on the superconducting properties of metal-oxide ceramics has been
studied very intensively.

A large number of theoretical models and mechanisms have been proposed to
explain this phenomenon and to explain the rich variety of magnetic and
superconducting properties of these materials. Besides the usual mechanism of
superconductivity based on the electron-\\
phonon interaction \cite{Bardeen} - \cite{Eliashberg}, mechanisms have been
proposed based on excitons \cite{Aleksandrov}, holes \cite{Hirch}, magnetic
interactions \cite{Anderson}, the interference between dielectric and
superconductor correlations \cite{Kopaev}, and so on.

Band-structure calculations \cite{Mattheis}-\cite{Krakauer} show that in
these compounds the Fermi surface can pass through points of high symmetry
corresponding to an electronic topological Lifshits transition \cite{Lifshits}.
In addition, in these compounds several of the energy bands overlap on the
Fermi surface \cite{Krakauer}, \cite{Herman}. For example, in
$YBa_{2}Cu_{3}O_{7-\delta}$, the number of bands overlapping on the Fermi
surfaces increases with the number of oxygen atoms \cite{Herman}: there
are two overlapping energy bands when $\delta=1$.  This suggests that
metal-oxide ceramics can be described using the multiband theory of
superconductivity \cite{Moskalenko} - \cite{Moskalenko_3}. An increase in the
number of energy bands on the Fermi surface leads to an increase in the total
density of electron states, and also to an additional interband
electron-electron interaction, which favors the superconducting state. This
interaction destroys the validity of the universal BCS relations, and the
thermodynamic characteristics depend significantly on the properties of the
anisotropic system.

An interesting feature of the two-band model is that the temperature of the
supercondu-\\
cting transition $T_{c}$ is independent of the sign of the interband
electron-electron interaction constant. Hence we can use the two-band model
for the usual electron-phonon mechanism of superconductivity, and also for the
mechanism based on repulsive interactions between carriers belonging to
different bands.

These features of the band structure explain the experimental behavior of the
thermody-\\
namic and electromagnetic characteristics of metal-oxide ceramics. In
particular, assuming moderate values of the coupling constants, the two-band
model predicts a high value of $T_{c}$, two energy gaps
$2\Delta_{1}/T_{c}>3.5$, $2\Delta_{2}/T_{c}<3.5$, a large negative value of
$d\ln T_{c}/d\ln V$ ($V$ is the volume), a positive curvature of the upper
critical field near the transition temperature, and so on \cite{Lee}-
\cite{Konsin}. These properties have been observed in a number of
experiments \cite{Verkin} - \cite{Welp}. It has been suggested \cite{Volovik}
that a gapless state in a pure two-band system can occur
for a fairly strong interband interaction and may explain the observed linear
temperature dependence of the heat capacity in high-temperature superconductors
in the low-temperature region \cite{Ishikawa}. In addition, the two-band
model can explain the decrease in $T_{c}$ with increasing oxygen disorder and
also with the replacement of copper atoms by a nonmagnetic impurity ($Al$,
$Zn$, and so on) \cite{Moskalenko_6},\cite{Moskalenko_7}.In Refs.
23 and 32 it was assumed that the basic mechanism responsible for the decrease
in $T_{c}$ is interband scattering of electrons by randomly distributed oxygen
vacancies or by impurities. Also the position of the Fermi level plays an
important role in determining $T_{c}$. Alloying or the introduction of oxygen
can change the position of the Fermi level and force it to pass through special
critical points of the energy spectrum, leading to an electronic topological
transition.  The dependence of $T_{c}$ on the position of the chemical
potential was considered in the two-band model in Ref.  25.

The purpose of the present review article is to consider the application of
the multiband model \cite{Moskalenko} to high-temperature superconductors
and to extend this theory to the case when points of high symmetry in momentum
space are located near the Fermi level.

The basic assumptions of the theory of superconductivity with overlapping
energy bands are discussed. A model Hamiltonian is formulated describing the
system and the main conclusions of the theory are presented. The differences
between the basic formulas of the microscopic theory of multiband
superconductors and the universal relations of the BCS-Bogolyubov theory are
pointed out. These differences depend on the properties of the material and
hence may explain the diversity of observed properties of superconductors,
including high-temperature superconductors. Since at the present time there is
no available theoretical approach capable of explaining the thermodynamic,
magnetic, and supercondu-\\
cting properties of high-temperature superconductors from a unified point of
view, it is useful to try to explain individual aspects of this many-sided
problem from the point of view of the multiband mechanism of superconductivity.

We note that current theories of band spectra in complex materials can
describe the band structure adequately, but not exactly. Hence the application
of the multiband theory to high-temperature superconductor compounds is to a
certain degree problematical. Never-\\
theless the existence of overlapping energy bands on the Fermi surface is
certain and so it is reasonable to use this theory to explain the properties
of high-temperature superconductors.

The two-band model is used in Sec. 2 to describe the thermodynamic properties
of $La_{2-x}(Ba,Sr)_{x}CuO_{4}$ (one wide band and one narrow band) and
$YBa_{2}Cu_{3}O_{7-\delta}$ (both bands with the same width).

The temperature dependence of the upper critical  field $H_{c2}$ in
high-temperature super-\\
conductors is considered in Sec. 3 with the help of the two-band model.

The effect of oxygen disorder and impurities replacing  $Cu$ in
$YBa_{2}Cu_{3}O_{7-\delta}$ is considered in Sec. 4. Interband scattering of
electrons by impurities turns out to be the basic mechanism leading to the
decrease of the transition temperature $T_{c}$ with increasing oxygen vacancies
or impurity atoms.

In Sec. 5 we consider the usual electron-phonon mechanism of superconductivity
in the two-band model with strong anisotropy and extremum points in $k$ space,
which lead to electronic topological transitions when oxygen atoms are
introduced into the system. The formation of flat regions on the Fermi surface
will be considered, as well as cylindrical cavities.

The dependence of $T_{c}$ on the filling of the energy bands is considered in
the model of three overlapping energy bands, assuming that superconductivity is
due only to the effective interband interaction of holes and that the sign of
the effective interaction does not affect $T_{c}$. Then this interaction can be
extended over a large part of $k$ space and the integrals in the equations
determining $T_{c}$ can be cut off at an energy of the order of electronic
energy.

\section{THERMODYNAMIC PROPERTIES}

The Hamiltonian describing a two-band system has the form \cite{Moskalenko}

\begin{eqnarray}
H&=&\sum_{n,\vec
k,\sigma}\varepsilon_{n}(\vec k)a_{n\vec k\sigma}^{\dagger}a_{n\vec k\sigma}
\nonumber\\ &-&\sum_{n,m,\vec k,\vec
k'}V_{nm}(\vec k,\vec k')a_{n\vec k\uparrow}^{\dagger}a_{n,-\vec k\downarrow}
^{\dagger}a_{m,-\vec k'\downarrow}a_{m,\vec k'\uparrow},\nonumber
\hspace{7.0cm}(2.1)
\end{eqnarray}
where $\varepsilon_n$ is the
energy of electrons in the $n$th band ($n,m=1,2$), $V_{nm}$ are the effective
matrix elements of the interband interaction, $a_{n\vec k\sigma}^{\dagger}$ and
$\bf a_{n\vec k\sigma}$ are the creation and annihilation operators of
electrons in the state $\vert n,\vec k, \sigma\rangle$.  Using (2.1), we obtain
a system of equations for the order parameters $\Delta_{1}$ and $\Delta_{2}$
$$
\hspace{4.8cm}\Delta_{n}=\frac{1}{2}\sum_{m}V_{nm}\sum_{\vec k}
\frac{th\left(\frac{\beta E_{m}(\vec k)}{2}\right)}{E_{m}(\vec k)},
\hspace{5.0cm}(2.2)
$$
where the energy of the elementary one-particle excitations has the form
$$
\hspace{5.4cm}E_{m}(\vec k)=\left(c_{m}^{2}(\vec k) +
\Delta_{m}^{2}\right)^{1/2}.  \hspace{5.2cm}(2.3)
$$

We will assume that the important mechanism in high temperature
superconductors is interband attractive interaction between quasiparticles.
Then we can put $V_{22}=0$ and $V_{12}, V_{11}\not = 0$ and the critical
transition temperature to the superconducting state $T_{c}$ has the form
$$
\hspace{6.8cm}T_{c}=\frac{2\gamma_{e}\omega_{D}}{\pi}e^{-\xi},
\hspace{5.89cm}(2.4)
$$
where
$$
\hspace{5.4cm}\xi=\frac{[\lambda_{12}^{2} + (\lambda_{11}^{2}/4)]^
{1/2}}{\lambda_{12}^{2}}-\frac{\lambda_{11}}{2\lambda_{12}^{2}},
\hspace{4.8cm}(2.5)
$$
$$\lambda_{11}=N_{1}V_{11},\ \lambda_{12}=(N_{1}N_{2})^{1/2}V_{12},$$
$N_{1}$ and $N_{2}$ are the densities of electronic states on the
corresponding parts of the Fermi surface. We see from (2.4) and (2.5) that
interband interactions between electrons lead to important
differences between these equations and the case of one-band superconductors
(the BCS theory). In the approximation of weak coupling we have for the order
parameters $\Delta_{1}$ and $\Delta_{2}$ at $T=0$
$$
\hspace{3.2cm}\Delta_{1}(0)=2\omega_{0}z^{1/(1 + nz^{2})}e^{-\xi}, \ \
\Delta_{2}(0)=2\omega_{0}z^{-1/[1 + (1/nz^{2})]}e^{-\xi}\hspace{2.3cm}(2.6)
$$
From (2.4) and (2.6) we have
$$
\hspace{3.5cm}
\frac{2\Delta_{1}(0)}{T_{c}}=\frac{2\pi}{\gamma_{e}}z^{1/(1+nz^{2})},
\frac{2\Delta_{2}(0)}{T_{c}}=\frac{2\pi}{\gamma_{e}}z^{-1/[1+(1/nz^{2})]},
\hspace{2.8cm}(2.7)
$$
where
$$
\hspace{4.3cm}
z=\frac{\Delta_{1}(0)}{\Delta_{2}(0)}=\frac{(\lambda_{11}/2) +
[(\lambda_{11}^{2}/4) + \lambda_{12}^{2}]^{1/2}}{\sqrt{n}\lambda_{12}}.
\hspace{3.72cm}(2.8)
$$

It follows from (2.7) that the ratio $2\Delta/T_{c}$ is not universal, but
depends on $z$ and on the ratio of the electronic densities of states
$n=N_{1}/N_{2}$.

The relative jump in the electronic heat capacity at the critical point
$T=T_{c}$ has the form
$$
\hspace{5.4cm}
(C_{s}^{e} - C_{n}^{e})/C_{n}^{e}\approx1,43A_{1}(z^{2}),\hspace{4.8cm}(2.9)
$$
where $C_{n}^{e}$ is the low-temperature electronic heat capacity of normal
metal, which in the two-band model has the form
$$
\hspace{6.0cm}
C_{n}^{e}=\frac{2\pi^{2}}{3}(N_{1} + N_{2})T.\hspace{5.4cm}(2.10)
$$
The quantity $A_{1}(z^{2})$ is given by \cite{Moskalenko_2},
\cite{Moskalenko_3}, \cite{Moskalenko_8}
$$
\hspace{5.7cm}
A_{1}(z^{2})=\frac{(n + z^{-2})^{2}}{(1 + n)(n + z^{-4})}.\hspace{4.58cm}
(2.11)
$$
Because of the factor $A_{1}(z^{2})$ in (2.9), the relative jump in the
electronic heat capacity in two-band superconductors is not universal. The
factor $A_{1}(z^{2})$ can be much less than unity and then the relative jump
in the electronic heat capacity will be small.

The thermodynamic critical field $H_{c}(T)$ in the low-temperature region
$(T/T_{c}\ll 1)$ and near $T_{c}$ is given by \cite{Moskalenko_2},
\cite{Moskalenko_3}, \cite{Moskalenko_8}
$$
\hspace{2.8cm}
\frac{H_{c}^{2}(T\sim 0)}{H_{c}^{2}(0)}\approx 1 -
2\chi\frac{T^{2}}{T_{c}^{2}}, \frac{H_{c}(T\to T_{c})}{H_{c}(0)}\approx
1.73A_{3}(z)\left(1 - \frac{T}{T_{c}}    \right),\hspace{1.29cm}(2.12)
$$
where
$$
\hspace{3.2cm}
H_{c}(0)=\left(4\pi\sum_{n}N_{n}\Delta_{n}^{2}(0)\right)^{1/2},\ \chi=1,06A_{3}
(z),\hspace{3.5cm}(2.13)
$$
$$A_{2}(z)=\frac{z(1+n)}{1+nz^{2}}\exp\left(-\frac{1-nz^{2}}{1+nz^{2}}\ln{z}
\right),
$$
$$
A_{3}(z)=\left(\frac{n + 1/z^{2}}{n+1/z^{4}}\right)z^{-1/(1 + nz^{2})}.
$$
Because of the functions $A_{2}(z)$ and $A_{3}(z)$ in (2.12), the
thermodynamic critical field is no longer universal, but depends significantly
on the properties of the anisotropic system.

One can attempt to use the above results for the thermodynamic properties of
two-band systems to describe the high-temperature superconductors
$La_{2-x}M_{x}CuO_{4}$ and $YBa_{2}Cu_{3}O_{x}$. In the first of these
compounds overlap of the two bands is possible: the first band consists of
copper $d_{x^2-y^2}$ and oxygen $2p$ hybridized orbitals and the
second band consists mainly of $d_{z^2}$ and oxygen $2p$ orbitals \cite{Lee}.
The second band is narrow and we have $N_{1}\ll N_{2}(n\ll 1)$.\\

TABLE I. Dependence of $T_c/1.14\omega_{0}$ on the two interaction parameters
$\lambda_{11}$ and $\lambda_{12}$.\\

\begin{tabular}{|c|c|c|c|c|c|}\hline
 $\lambda_{12}$ & 0.1 & 0.2 & 0.3   & 0.4   & 0.5 \\ \hline
$\lambda_{11}=0.1$ & 0.002 & 0.02 & 0.059 & 0.11  & 0.164 \\
 0.3               & 0.048 & 0.082 & 0.127 & 0.177 & 0.226\\
 0.5               & 0.146 & 0.173 & 0.210 & 0.25  & 0.291\\ \hline
\end{tabular}\\
\\

In the case of $YBa_{2}Cu_{3}O_{x}$, the number of bands on the Fermi surface
depends on the value of $x$ (Ref. 15). Since the bands have about the same
width we can put $N_{1}\approx N_{2}$ and $n\approx 1$ in this case. Numerical
estimates of the thermodynamic characteristics using the formulas given above
are listed in Tables I-III. The observed value $T_{c}\approx 40^{\circ}K$ in
$La_{2-x}M_{x}CuO_{4}$ can be obtained for small, physically reasonable values
of the frequency $\omega_{0}$. For example, the last column of Table I for
$T_{c}=40^{\circ}K$ corresponds to $\omega_{0}=214, 155$, and
$120^{\circ}K$. For the metallic ceramics $YBa_{2}Cu_{3}O_{7-\delta}$ we
have $\omega_{0}\approx 863^{\circ}K$ and it is not difficult to obtain the
observed value $T_{c}\approx 100 ^{\circ}K$ using the theory presented
above. The possible values of $T_{c}$ for this case are given in the last
column Table III. Tables II and III show that in the weak coupling
approximation $2\Delta_{1}/T_{c}$ and $2\Delta_{2}/T_{c}$ can be different,
the jump in the electronic heat capacity can be small, and $H_{c}(T)$ can
differ significantly from the prediction of the BCS theory. These features are
observed in high-temperature superconductors. In particular, in a series of
tunneling experiments (see Ref. 26, for example) two energy gaps were observed
in $La_{2-x}(Ba,Sr)_{x}CuO_{4}$, where
$$2\Delta_{1}/T_{c}>3,5\ and\ 2\Delta_{2}/T_{c}<3,5.$$

Small values of $2\Delta_{2} /T_{c}$ have also been obtained in infrared
absorption experiments (see Ref. 27 and the bibliography in Ref. 20).

The volume dependence of $T_{ c}$ predicted by this model is of interest. We
obtain from (2.4) and (2.5)
\begin{eqnarray}
\frac{d\ln T_{c}}{d\ln V}
&=&\frac{d\ln \omega_{o}}{d\ln V} -
\left(\frac{\lambda_{11}}{4\lambda_{12}^{2}\eta} - \frac{\lambda_{11}}
{2\lambda_{12}^{2}} \right)\frac{d\ln \lambda_{11}}{d\ln V} \nonumber\\
&-&\left[\frac{1}{\eta} - \frac{2}{\lambda_{12}}\left(\eta -
\frac{\lambda_{11}}{2}\right)\right]\frac{d\ln \lambda_{12}}{d \ln V},
\nonumber\hspace{7.5cm}(2.14)
\end{eqnarray}
where $\eta=(\lambda_{12}^{2} + 1/4\lambda_{11}^{2})^{1/2}$. The dependence of
$d\ln T_{c}/d\ln V$ on $T_{c}$ is shown in Fig. 1 for
$$
\frac{d\ln\omega_{0}}{d\ln V}=-1,\ \frac{d\ln \lambda_{11}}{d\ln V}=\frac{d\ln
\lambda_{12}}{d\ln V}=-2
$$
and for $\omega_{0}=181, 360, 500^{\circ}K$ (curves 1-3, respectively). It
follows from Fig. 1 that $d\ln T_{c}/d\ln V$ can be large and negative, as
observed experimentally in $La_{2-x}M_{x}CuO_{4}$ compounds \cite{Grissen}.
It is not necessary to assume anomalously large values of the derivatives with
respect to volume of the parameters of the theory $\omega_{0}$, $\lambda_{11}$,
$\lambda_{12}$; physically reasonable values can be used.

We note that the behavior of the quantity $d\ln T_{c}/d \ln V$ as a function of
$T_{c}$ in the two-band model is similar to the behavior of this quantity in
the theory of two-dimensional superconductivity \cite{Grissen}, \cite{Lable},
in which the logarithmic singularity in the electronic density of states is
taken into account.\\ 

TABLE II. The case $n=0.1$ in $La_{2-x}(Ba,Sr)_{x}CuO_{4}$.\\

\begin{tabular}{|c|c|c|c|c|c|c|c|}\hline
$\lambda_{11}$ & $\lambda_{12}$  & $z$  & $\frac{2\Delta_{1}}{T_{c}}$

& $\frac{2\Delta_2}{T_c}$ & $\frac{C_s^e - C_n^e}{C_n^e}$ & $\chi$ & $A_3(z)$
 \\ \hline
0.1 & 0.3 & 3.73 & 6.07 & 1.63 & 0.365 & 2.58 & 0.734  \\
0.1 & 0.4 & 3.58 & 6.12 & 1.71 & 0.388 & 2.42 & 0.741  \\
0.1 & 0.5 & 3.49 & 6.15 & 1.76 & 0.403 & 2.33 & 0.743   \\
0.3 & 0.2 & 6.32 & 5.06 & 0.80 & 0.202 & 5.28 & 0.771   \\
0.3 & 0.3 & 5.12 & 5.50 & 1.07 & 0.171 & 4.08 & 0.743   \\
0.3 & 0.5 & 4.25 & 5.86 & 1.38 & 0.304 & 2.96 & 0.733 \\ \hline
\end{tabular}\\
\\

Therefore the features of the density of electron states such as overlapping
energy bands and electronic topological transitions can be crucial to the
thermodynamic properties of high-temperature superconductors. It may be
necessary to take into account these features simultaneously.

\section{UPPER CRITICAL FIELD}

The properties of two-band superconductors near the upper critical field can
be studied using the generalized Ginzburg-Landau equations
\cite{Moskalenko_9}-\cite{Tilley}.

Putting $V_{22}=0$, the expression for $H_{c2}$ in the low-temperature region
($T \to 0$) reduces to the form \cite{Palistrant}
$$
\hspace{3.5cm}
\frac{H_{c2}(T \to 0)}{H_{c2}(0)}=1 +
\frac{16\gamma_{e}}{\pi^2e_{0}^{2}}\left(\frac{T}{T_{c}}
\right)^{2}\Phi(T)\exp(\nu(\lambda) - \nu(1)),\hspace{2.4cm}(3.1)
$$

$$
\hspace{5.0cm}
H_{c2}(0)=\frac{\pi^{2}T_{c}^{2}e_{0}^{2}}{2\gamma_{e}e\theta_{1}\theta_{2}}
\exp(\nu(1) - \nu(\lambda)),\hspace{4.1cm}(3.2)
$$
\begin{eqnarray}
&\Phi(T)&\nonumber\\
&=&(\lambda\gamma^{-} + \frac{1}{\lambda}\gamma^{+})[\xi'(2) +
\frac{1}{2}\xi(2)(2\ln\frac{T}{T_c}\nonumber\\
&+&2\ln\frac{4}{\pi e_0} + \nu(\lambda) - \nu(1) + \ln\lambda)],\nonumber
\end{eqnarray}
$$
\nu(\lambda)=\left[ \left(\ln \lambda + \frac{\lambda_{11}}{\lambda_{12}^{2}}
\right)^{2} + \frac{4}{\lambda_{12}^{2}} \right]^{1/2},
$$
$$
\hspace{3.0cm}
\gamma^{\pm}=\frac{1}{2}\left[1 \pm \left(\frac{\ln \lambda +
\lambda_{11}\lambda_{12}^{-2}}{\nu(\lambda)}\right)  \right];
\hspace{6.75cm}(3.3)
$$
where $\lambda=v_1/v_2$, with $v_1$ and $v_2$ being the velocities of electrons
in the corresponding cavity of the Fermi surface, $e_{0}$ is the base of
natural logarithms, and $e$ is the charge of the electron.\\

TABLE III. The case $n=1$ in $YBa_{2}Cu_3O_x$.\\

\begin{tabular}{|c|c|c|c|c|c|c|c|c|}\hline
$\lambda_{11}$ & $\lambda_{12}$  & $z$  & $\frac{2\Delta_{1}}{T_{c}}$

& $\frac{2\Delta_2}{T_c}$ & $\frac{C_s^e - C_n^e}{C_n^e}$ & $\chi$ & $A_3(z)$
& $T_c,K$ (for $\omega_0=863$ $K$)  \\ \hline
0.1 & 0.3 & 1.18 & 3.75 & 3.18 & 1.39  & 0.76 & 0.994 & 58  \\
0.2 & 0.3 & 1.39 & 3.91 & 2.82 & 1.30  & 0.89 & 0.978 & 89\\
0.4 & 0.1 & 4.24 & 3.78 & 0.89 & 0.794 & 1.77 & 0.951 & 93 \\
0.4 & 0.3 & 1.87 & 4.02 & 2.15 & 1.09  & 1.15 & 0.949 & 165 \\
0.5 & 0.5 & 1.62 & 4.0  & 2.47 & 1.19  & 1.02 & 0.961 & 286 \\ \hline
\end{tabular}\\
\\

For temperatures close to the superconducting transition temperature $T_{c}$
we obtain \cite{Moskalenko_2}, \cite{Palistrant}
\begin{eqnarray}
\frac{H_{c2}(T\sim T_{c})}{H_{c2}(0)}&=&\frac{8\gamma_ev_1v_2}
{e_{0}^{2}[v_1^{2}\eta_{1} + v_2^{2}\eta_{2}]}\exp(\nu(\lambda)
- \nu(1))\frac{6}{7\xi(3)}\nonumber\\
&\times&\left(1 - \frac{T}{T_c}  \right)\nonumber\\
\times& &\left\{ 1 + \left(1- \frac{T}{T_c}\right)\left[\frac{\frac
{v_1^2}{v_2^2}\eta_{1} +
\frac{v_2^2}{v_1^2}\eta_2}{\left(\frac{v_1}{v_2}\eta_1 +
\frac{v_2}{v_1}\eta_2\right)^2}\frac{3I}{10}\xi(5)\left(\frac{6}
{7\xi(3)} \right)^2 - \frac{3}{2}\right]\right\},\nonumber\hspace{1.8cm}(3.4)
\end{eqnarray}
where
$$
\hspace{5.57cm}
\eta_{1,2}=(1 \pm \eta)/2,\ \eta=\lambda_{11}/\lambda_{12}^{2}\nu(1).
\hspace{3.9cm}(3.5)
$$
Putting $v_1=v_2$ in (3.1)-(3.4), we obtain the corresponding
relations for an ordinary one-band superconductor \cite{Gor'kov}, \cite{Maki}:
$$
\hspace{3.0cm}
\frac{H_{c2}(T \to 0)}{H_{c2}(0)}=1 + \frac{16\gamma_e}{\pi^2 e_0^2}\left(
\frac{T}{T_c}\right)^2\left\{\xi(2)\ln\frac{T}{T_c} + \xi'(2) +
\xi(2)\ln\frac{4}{\pi e_0} \right\},\hspace{0.5cm}(3.6)
$$
\begin{eqnarray}
\frac{H_{c2}(T \to T_c)}{H_{c2}(o)}&=&\frac{8\gamma_e}{e_0^2}\frac{6}{7\xi(3)}
\left(1 - \frac{T}{T_c}    \right)\Bigg\{1 + \left(1 - \frac{T}{T_c}\right)
\nonumber\\
& &\times\left[\frac{31}{10}\xi(5)\left(\frac{5}{7\xi(3)}\right)^2 - \frac{3}
{2}\right]\Bigg\}\nonumber\hspace{6.89cm}(3.7)
\end{eqnarray}\\

Letting $H_{c2}^0(0)$ and $T_{c 0}$ be the upper critical field and critical
temperature of a low-tem-\\
perature one-band superconductor, we obtain from (3.2)
$$
\hspace{4.5cm}
H_{c2}(0)/H_{c2}^0(0)=(T_c/T_{c0})^2\frac{v_1}{v_2}\exp(\nu(1) -
\nu(\lambda)).\hspace{2.8cm}(3.8)
$$
Numerical estimates based on (3.8) show that the upper critical field for
two-band supercon-\\
ductors at $T=0$ can be two or three orders of magnitude larger than
$H_{c2}(0)$ for ordinary superconductors. The quantity $H_{c2}(0)$ is large
because of the high value of $T_c$ and the fact that $v_1/ v_2 > 1$
or $\gg 1$.

Figure 2 shows the dependence $H_{c2}(T)/H_{c2}(0)$ obtained from (3.1) and
(3.4) for $T\sim 0$ and $T\sim T_c$, respectively and the extrapolation of
these functions. We see that the curvature of this dependence changes as
$v_1/v_2$ increases. The curvature of curves 3 and 4 is observed
experimentally. We see that if the second band contains heavy carriers (small
velocity on the Fermi surface) then the two-band model qualitatively describes
the behavior of $H_{c2}$ as a function of temperature.

\section{EFFECT OF OXYGEN VACANCIES OR
\newline IMPURITIES ON THE TRANSITION
\newline TEMPERATURE IN HIGH-TEMPERATURE
\newline SUPERCONDUCTORS}

Here the two-band model is used to attempt to understand the experimentally
observed decrease in $T_c$ with a decrease in the concentration of oxygen or
with the disordering of oxygen vacancies in $YBa_2Cu_3O_{7 - \delta}$.
Disordering appears here as a nonmagnetic decoupling factor of
superconducting pairs. We start from the Hamiltonian describing two-band
superconductors
$$
\hspace{6.6cm}H=H_0 + H_{1},\hspace{6.2cm}(4.1)
$$
where $H_{0}$ is the Hamiltonian (2.1) of a pure two-band material and $H_{1}$
has the form
$$
\hspace{3.9cm}H_{1}=\frac{1}{V}\sum_{\sigma}\sum_{n,\vec k}\sum_{n',\vec k'}
a_{n\vec k\sigma}^{\dagger}a_{n'\vec k'\sigma}\rho(\vec k-\vec k')
U_{nn'}(\vec k-\vec k'),\hspace{2.4cm}(4.2)
$$
$$
\hspace{4.8cm}
\rho(\vec k-\vec k')=\sum_{j}\exp\left[-i(\vec k - \vec k')\vec r_{j}\right].
\hspace{4.35cm}(4.3)
$$

Hence we assume that the metallic phase of $YBa_2Cu_3O_7$ with the maximum
supercondu-\\
cting transition temperature $T_c\approx 90^{\circ} K$ corresponds
to the two-band Hamiltonian $H_0$. A decrease in the number of oxygen atoms
(increase in $\delta$) leads to a disordering of the system, and the
additional term (4.2) in the Hamiltonian describes the interaction of
electrons with lattice defects ($U_{nn'}$ is the potential energy describing
the scattering of electrons by oxygen vacancies). The sum in (4.3) is taken
over the randomly distributed oxygen vacancies.

Hence the effect of the oxygen composition (or disordering) on $T_c$ reduces
formally to the problem of determining the effect of a nonmagnetic impurity on
the superconducting transition temperature in two-band superconductors
\cite{Moskalenko_9}, \cite{Moskalenko_10}. A decrease in the number of oxygen
atoms (increase in $\delta$) corresponds to an increase in the concentration of
impurities. The critical temperature of the two-band superconductor is found as
the eigenvalue of the equation for a bound state of a pair of electrons or
holes with zero binding energy. This equation is obtained starting from the
Dyson equation for the two-particle Green's function of the system (4.1).
Averaging it over the randomly distributed oxygen vacancies, we obtain a
linearized system of equations for the superconductor order parameters
$\Delta_1$ and $\Delta_2$ (Refs. 41 and 41):
\begin{eqnarray}
\Delta_{1}&=&\left[\lambda_{11}\xi -
(\lambda_{12} - jn^{1/2}\lambda_{12})\left(1 +
\frac{\tau_{12}}{\tau_{21}}\right)^{-1}J_2(\tau_{12})\right]\Delta_{1}
\nonumber\\
&+&\left[\lambda_{12}\frac{1}{\sqrt{n}}\xi + (\lambda_{11} -
jn^{1/2}\lambda_{12})\left(1 +
\frac{\tau_{12}}{\tau_{21}}\right)^{-1}J_2(\tau_{12})\right]\Delta_{2},
\nonumber\\
\Delta_{2}&=&[\xi n^{1/2}\lambda_{12}-\lambda_{12}n^{1/2}J_2(\tau_{12})(1+\frac{\tau_{12}}
{\tau_{21}})^{-1}]\Delta_1\nonumber\\
&+&n^{1/2}\lambda_{12}J_2(\tau_{12})(1+\frac{\tau_{12}}{\tau_{21}})^{-1}
\Delta_2,\nonumber\hspace{8.28cm}(4.4)
\end{eqnarray}
where
$$
\lambda_{11}=N_1V_{11}\ \lambda_{12}=(N_1N_2)^{1/2}V_{12},\
j=\frac{\tau_{12}}{\tau_{21}}\frac{N_2}{N_1}=\frac{\tau_{12}}{\tau_{21}}
\frac{1}{n},
$$
$$
\hspace{5.7cm}
\xi=\ln\frac{2\gamma\omega}{\pi T},\ n=\frac{N_1}{N_2}
\hspace{5.7cm}(4.5)
$$
\begin{eqnarray}
J_{2}&=&\int_{0}^{\infty}\frac{th\left(\frac{\beta
y}{4}\right)\left(\tau_{12}^{-1} + \tau_{21}^{-1}\right)dy}{y(y^2 +
1)}\nonumber\\ &=&\Psi\left(\frac{1}{2} +
\frac{\beta}{4\pi}\left(\frac{1}{\tau_{12}} +
\frac{1}{\tau_{21}}\right)\right) - \Psi\left(\frac{1}{2}\right),
\nonumber\hspace{7.5cm}(4.6)
\end{eqnarray}
$N_1$ and $N_2$ are the densities of electron states in the corresponding
cavity of the Fermi surface, $V_{11}$ and $V_{12}$ are the effective
electron-electron intraband and interband interactions, and $\tau_{12}$ is
the relaxation time of interband scattering by oxygen vacancies.

In obtaining (4.4) we assumed that the important mechanism for
high-temperature superconductivity is interband attractive interactions
between electrons. Then we can put $V_{22}\ll V_{12}$ and neglect
electron-electron interactions within the second band.

The condition for a nontrivial solution of (4.4) determines $\xi$ and hence
$T_{c}$ and has the form
$$
\hspace{6.2cm}
a\xi_c^2 - b\xi_c + c=0,\hspace{6.0cm}(4.7)
$$
where
$$
\hspace{3.9cm}a=-\lambda_{12}^{2},\ \xi_c=\ln \frac{2\omega_0\gamma}{\pi T_c},\
b=\lambda_{11} - \lambda_{12}^2J_2(\tau_{12}),\hspace{3.2cm}(4.8)
$$
\begin{eqnarray}
c&=&1 + \lambda_{11}\left(1 + \frac{\tau_{12}}{\tau_{21}}\right)^{-1}
J_2(\tau_{12})\nonumber\\
&-&\lambda_{12}n^{1/2}(1 + j)\left(1 + \frac{\tau_{12}}{\tau_{21}}\right)^{-1}
J_2(\tau_{12}).\nonumber
\end{eqnarray}
Of the two solutions of (4.7), we choose the physically reasonable solution
satisfying the condition $T_c\to 0$ when $\lambda_{11}$, $\lambda_{12}\to 0$.
This solution has the form
$$
\hspace{3.0cm}\ln \frac{T_c}{T_{c0}}=-\alpha\left[\Psi\left(\frac{1}{2} +
\frac{\beta}{4\pi}\left(\frac{1}{\tau_{12}} + \frac{1}{\tau_{21}}\right
)\right) - \Psi\left(\frac{1}{2}\right)\right],\hspace{3.0cm}(4.9)
$$
where
$$\hspace{3.0cm}
\alpha=\frac{1}{2}\left\{1 - \frac{\lambda_{11} - 2[\lambda_{11} -
\lambda_{12}n^{1/2}(1 + j)](1 + jn)^{-1}}{(\lambda_{11}^{2} +
4\lambda_{12}^{2})^{1/2}}\right\}.\hspace{2.4cm}(4.10)
$$

It follows from (4.9) that the critical temperature $T_c$ decreases with
increasing concentra-\\
tion of oxygen vacancies (an increase in the factor $1/\tau_{12} +
1/\tau_{21}$). The rate of decrease depends very strongly on the coefficient
$\alpha$ given by (4.10). In turn, $\alpha$ depends on the electron-electron
interaction constants $\lambda_{11}$ and $\lambda_{12}$, on the ratio
$N_1/N_2$, and on the parameter $j$ determining the difference between the
cavities in the Fermi surface ($j=1$ for spherically symmetric cavities).
Since $\alpha < 1$, the decrease in $T_c$ is not as sharp as in the case of a
superconductor with a paramagnetic impurity \cite{Abrikosov}.

For low concentrations of oxygen vacancies
$$
\frac{\beta_c}{2}\left(\frac{1}{\tau_{12}} + \frac{1}{\tau_{21}}\right) \ll 1
$$
(4.9) reduces to
$$
\hspace{6.0cm}
T_c=T_{c0} - \frac{\pi\alpha}{8\tau_{12}}(1 + nj).\hspace{5.0cm}(4.11)
$$
The temperature $T_c$ decreases linearly with increasing $\tau_{12}^{-1}$.

In the other limiting case
$$
\frac{\beta_c}{2}\left(\frac{1}{\tau_{12}} + \frac{1}{\tau_{21}}\right) \gg 1
$$
we have
$$
\hspace{4.3cm}T_c=T_{c0}^{1/(1-a)}\left[\frac{\gamma}{\pi\tau_{12}}(1 +
jn)\right]^{\alpha/(\alpha - 1)}\hspace{4.5cm}(4.12)
$$
In this expression $T_c\to 0$ only when the concentration of impurities
becomes infinite. This means that there is no critical concentration of
impurities for which $T_c=0$. Hence interband scattering of electrons by
oxygen vacancies $YBa_{2}Cu_{3}O_{7-\delta}$ significantly suppresses
superconductivity. The dependence of $T_{c}/T_{c0}$ on $1/\tau_{12}T_{c0}$ is
shown in Fig. 3 for different values of the parameters of the theory.

We next consider the densities of electron states of disordered two-band
superconductors. The ratio $N_n(\omega)/N_n$ ($n=1,2$) is given by
\cite{Moskalenko_11}, \cite{Moskalenko_12}
$$
\hspace{3.75cm}
\frac{N_n(\omega)}{N_n}=Im\frac{u_n(\omega)}{(1 - u_n^2(\omega))^{1/2}}=Re
\frac{u_n(\omega)}{(u_n^2(\omega )- 1)^{1/2}},\hspace{2.8cm}(4.13)
$$
where $u_1(\omega)$ and $u_2(\omega)$ satisfy the system of equations
\begin{eqnarray}
\frac{\omega}{\Delta_1}&=&u_1(\omega) + \alpha_1\frac{u_1(\omega) -
u_2(\omega)}{\left(1 - u_2^2(\omega)\right)^{1/2}},\
\alpha_1=\frac{1}{2\tau_{12}\Delta_1},\hspace{6.48cm}(4.14)\nonumber\\
\frac{\omega}{\Delta_2}&=&u_2(\omega)
+ \alpha_2\frac{u_2(\omega) - u_1(\omega)}{\left(1 -
u_1^2(\omega)\right)^{1/2}},\alpha_2=\frac{1}{2\tau_{12}\Delta_2}.\nonumber
\end{eqnarray}
For completeness we note that the order parameters $\Delta_1$ and $\Delta_2$
are determined from the following system of equations at arbitrary
temperature:
$$
\Delta_n=\frac{\pi}{\beta}\sum_m V_{nm}N_m\sum_m\frac{1}
{\left(u_m^2(\Omega) + 1  \right)^{1/2}},\  \Omega=(2n + 1)\frac{\pi}{\beta},
$$
\begin{eqnarray}
\frac{\Omega}{\Delta_1}&=&u_1(\Omega) + \alpha_1(\Omega)\frac{u_1(\Omega) -
u_2(\Omega)}{\left(1 + u_2^2(\Omega)\right)^{1/2}}\frac{\omega}{\Delta_2}
\nonumber\hspace{7.79cm}(4.15)\\
&=&u_2(\Omega) + \alpha_2(\Omega)\frac{u_2(\Omega) - u_1(\Omega)}{\left(1 +
u_1^2(\Omega)\right)^{1/2}}.\nonumber
\end{eqnarray}

It is not difficult to show that the density of electron states and also the
order parameters $\Delta_1$ and $\Delta_2$ of the superconducting phase depend
significantly on the relaxations times of interband scattering of electrons by
oxygen vacancies. An analytical method of calculating the density of electron
states has been given by one of the authors \cite{Moskalenko_12}. It
yields a nonanalytic dependence of the density of states on the small
concentration of impurities (oxygen vacancies) in the frequency regions near
the order parameters $\Delta_n$ and uses an expansion in the small parameters
of the theory in the region far from the order parameters. Numerical
calculations of the densities of states based on (4.13) for different
values of the parameters of the theory were also discussed in Ref. 45.

The energy gap of a two-band superconductor corresponds to the highest
frequency at which the density of states still equal to zero. In spite of the
fact that a pure two-band superconductor has two energy gaps, the addition of
impurities (disorder) leads to a single energy gap in the two-band model
because of mixing of states belonging to different cavities of the Fermi
surface by the impurity (oxygen vacancies). A proof of this statement is the
fact that both densities of electron states become nonzero
simultaneously \cite{Moskalenko_12}. The decrease in the number of
energy bands with decreasing number of oxygen atoms in
$YBa_{2}Cu_{3}O_{7-\delta}$ is an interesting question and can be understood
assuming that oxygen vacancies lead to a decoupling because of interband
scattering of carriers by the vacancies.  The result is a single hybridized
band instead of two overlapping bands. We note that the above theory can also
be used to describe the behavior of $T_c$ as a function of the concentration of
impurities $x$ in $YBa_2(Cu_{1-x}M_x)_3O_{7-\delta}$, where $M=Al, Mo, Zn$. In
this case $\tau_{12}$ is determined by interband scattering of conduction
electrons from the randomly distributed impurities. The dependence of $T_c$ on
$1/\tau_{12}$ obtained in Fig. 3 is qualitatively consistent with the observed
dependence of $T_c$ on the concentration of impurities \cite{Takabatake}
(Fig.  4).

\section{TWO-BAND MODEL AND ELECTRONIC
\newline TOPOLOGICAL TRANSITIONS}

Recently an interesting feature has been observed in the experimental
dependence of the superconducting transition temperature of
$YBa_{2}Cu_{3}O_{7-\delta}$ as a function of $\delta$ and also in
$(La_{1-x}Sr_{x})_{2}CuO_{4}$ as a function of $x$. The transition temperature
$T_{c}$ is practically constant (step-like) in the region $0.2\le \delta
\le 0.6$ in $YBa_{2}Cu_{3}O_{7-\delta}$ (Ref. 47) and in the region
$0.05\le x \le 0.08$ in $(La_{1-x}Sr_{x})_{2}CuO_{4}$ (Ref. 48).

In Secs. 5 and 6 we use the theory of superconductors with overlapping energy
bands and electronic topological transitions to obtain the dependence of
$T_{c}$ on the position of the chemical potential $\mu$ and to understand the
origin of the step-like behavior of $T_{c}$.

In the present Section we consider the dependence of the superconducting
transition temperature in the two-band model assuming the phonon mechanism of
superconductivity and taking into account the Van Hove-Lifshits features in
the electronic energy spectrum. The case of a topological transition in
two-band superconductors with the formation of elliptical cavities in the
Fermi surface was considered in Refs. 49 and 50. Here we will consider
strongly anisotropic (quasi-one-dimensional and quasi-two-dimensional)
three-dimensional systems. The presence of singular points in momentum space
of anisotropic systems of this kind leads to electronic topological transitions
accompanied by the formation of flat areas on the Fermi surface or cylindrical
cavities when the Fermi level passes through special critical points
$\varepsilon_{kn}$.

The formation of flat areas on the Fermi surface in a one-band system has been
studied in a number of papers \cite{Palistrant_4}-\cite{Palistrant_6} and has
been confirmed by  numerous experiments in the intermetallic compound
$AuGa_{2}$.

The electronic topological transitions considered here can be observed in
high-temperature superconductors where the atoms are clearly arranged in
chains (one-dimensional) or in a plane (two-dimensional). The possibility of
topological transitions in high-temperature ceramics was discussed in Refs. 30
and 54. We start from the Hamiltonian (2.1) with $V_{nm}\not =0$ ($n,m=1,2$),
which corresponds to including intraband and interband interactions.

1. We consider first the formation of flat areas on the Fermi surface. We
assume a strongly anisotropic system where the trajectories of the electrons
are nearly one-dimensional in a certain region of momentum space. Near the
critical points the energy of an electron in the $n$th band can be written in
the form
$$
\hspace{6.2cm}\varepsilon_{n}(p)=\varepsilon_{kn} + \frac{p_{z}^{2}}{2m_n},
\hspace{6.3cm}(5.1)
$$
where $m_n$ is the effective mass of the electron.

With the help of the modified Hamiltonian (2.1) and the dispersion law (5.1),
we obtain the following system of equations for the order parameters
$\Delta_n$ near the transition temperature ($T\approx T_{c}$):
\begin{eqnarray}
\Delta_{1}&=&A_1V_{11}I_{1}(\beta, \varepsilon'_{k_1})\Delta_1 +
A_2V_{21}I_{2}(\beta, \varepsilon'_{k_2})\Delta_2,\nonumber
\hspace{7.0cm}(5.2)\\
\Delta_{2}&=&A_1V_{12}I_{1}(\beta, \varepsilon'_{k_1})\Delta_1 +
A_2V_{22}I_{2}(\beta, \varepsilon'_{k_2})\Delta_2\nonumber,
\end{eqnarray}
where
$$
\hspace{3.79cm}
I_{n}(\beta, \varepsilon'_ {kn})=\int_{0}^{\omega_D}\frac{d\varepsilon}
{(\varepsilon + \varepsilon'_{kn})^{1/2}}\Theta(\varepsilon +
\varepsilon'_{kn})\frac{th\left(\frac{\beta\varepsilon}{2}\right)}
{\varepsilon},\hspace{3.2cm}(5.3)
$$
$$
\varepsilon'_{kn}=\mu - \varepsilon_{kn},\ \ A_{n}=N_{n}\varepsilon_{F}^{1/2}\alpha_{n},
$$
$$
\hspace{4.8cm}
N_{n}=\frac{m_{n}p_{Fn}}{2\pi^{2}}, \alpha_{n}=\frac{2}{\pi}\frac{\left(
\int\int dp_{x}dp_{y}\right)_{n}}{4p_{Fn}^{2}}.\hspace{4.3cm}(5.4)
$$
The transition temperature $T_{c}$ is obtained from the condition that the
determinant of the system (5.2) must vanish:
\begin{eqnarray}
& &\lambda_{11}\tilde \varepsilon_F^{1/2}\alpha_{1}I_{1}(\tilde \beta_{c},
\tilde \varepsilon'_{k1}) +
\lambda_{22}\tilde\varepsilon_F^{1/2}\alpha_{2}I_{2}(\tilde\beta_c,\tilde
\varepsilon_{k2})\nonumber\\
& &-(\lambda_{11}\lambda_{22}-\lambda_{12}\lambda_{21})I_1(\tilde\beta_c,
\tilde\varepsilon'_{k1})I_{2}(\tilde \beta_{c},\tilde\varepsilon'_{k2})
\tilde\varepsilon_{F}\alpha_{1}\alpha_{2} - 1=0,\nonumber\hspace{4.8cm}(5.5)
\end{eqnarray}
where
$$
\hspace{3.0cm}
I_{n}(\beta_c, \varepsilon'_{kn})=\int_{0}^{\tilde\omega_D}\frac{dx}{\left(x +
\tilde\varepsilon'_{kn}\right)^{1/2}}\Theta(x +
\tilde\varepsilon'_{kn})\frac{th\left(\frac{\beta_cx}{2}\right)}{x},
\hspace{3.65cm}(5.6)
$$
$$
\hspace{0.2cm}
\beta_{c}=\frac{T_0}{T_c},\ \tilde\varepsilon'_{kn}=\frac{\varepsilon'_{kn}}{2T_0},\
\tilde\varepsilon_{F}=\frac{\varepsilon_{F}}{2T_0},
$$
$$
\tilde\omega_{Dn}=\frac{\omega_{Dn}}{2T_0},\ \lambda_{nm}=(N_nN_m)^{1/2}V_{nm},
$$
$T_0$ is an arbitrary temperature introduced to obtain dimensionless quantities
in (5.5).

Figure 5 shows the dependence of the superconducting transition temperature
$T_{c}$ on the parameter $\tilde \varepsilon'_{k1}=\tilde\mu -
\tilde\varepsilon_{k1}$, which varies because of the introduction of oxygen
(change in the concentration of impurities) or because of pressure. These
results were obtained from the solution of (5.5).

In Fig. 5a note that $T_{c}$ varies only slightly in the region
$-12<\tilde\varepsilon'_{k1}<-10$. Therefore the topological transitions
considered here can lead to a step-like dependence of $T_{c}$ on the
concentration of oxygen in $YBa_{2}Cu_{3}O_{7-\delta}$ (Fig. 6) or on the
concentration of impurities in $(La_{1-x}Sr_{x})_{2}CuO_{4}$ (Ref. 48).
Topological transitions accompanied by the formation of flat areas on the
Fermi surface can lead to a double maximum in the dependence of $T_{c}$ on
$\tilde\varepsilon'_{k1}$ (Fig. 5b). Scattering by impurities would lead to a
broadening of these maxima \cite{Palistrant_4}-\cite{Palistrant_6}.
We note that this form of the dependence of $T_{c}$ on pressure was observed in
$Bi_{2}Sr_{2}CaCu_{2}O_{8}$ (Ref. 55).

2.  We next consider electronic topological transitions accompanied by the
formation of flat areas on the Fermi surface (because of one-dimensional
motion of electrons in the first band) and the formation of a cylindrical
cavity in the Fermi surface (because of two-dimensional motion of electrons in
the second band).

The dispersion laws near the critical points ($\nabla\varepsilon_{m}=0$) are
written in the form
$$
\hspace{6.0cm}\varepsilon_1(p)=\varepsilon_{k1} +
\frac{p_z^2}{2m_1},\hspace{6.0cm}(5.7)
$$
$$
\varepsilon_2(p)=\varepsilon_{k2} + \frac{1}{2m_2}(p_x^2 + p_y^2).
$$
In this case the temperature of the superconducting transition is determined
from the equation
\begin{eqnarray}
& &\lambda_{11}\tilde\varepsilon_{F}^{1/2}\alpha_{1}I_{1}(\tilde\beta_{c},
\tilde\varepsilon'_{k1}) +  \lambda_{22}I_{2}'(\beta_{c},\tilde
\varepsilon'_{k2})\nonumber\\
& &-(\lambda_{11}\lambda_{22} -
\lambda_{12}\lambda_{21})\tilde\varepsilon_{F}^{1/2}\alpha_{1}I_{1}(\beta_c,
\tilde\varepsilon'_{k1})I_{2}'(\tilde\beta_c, \tilde\varepsilon'_{k2}) - 1=0,
\nonumber\hspace{5.0cm}(5.8)
\end{eqnarray}
where
$$
\hspace{3.5cm}
N_{2}=\frac{m_2(\int dp_x)_2}{2\pi^2},\ I_{2}'(\tilde\beta_c,\tilde
\varepsilon'_{k2})=\int_{\tilde\varepsilon'_{k2}}^{\tilde
\omega_D}dx\frac{th\left(\frac{\beta_c x}{2}\right)}{x}.\hspace{3.5cm}(5.9)
$$

The dependence of $T_c$ on $\tilde\varepsilon'_{k1}=\tilde\mu -
\tilde\varepsilon_{k1}$ obtained by solving (5.8) is shown in Fig. 7 for the
case
$$
\lambda_{11}=0.25,\ \lambda_{12}=0.1,\ \lambda_{22}=0.3,\ \alpha_{1}=0.3,
$$
$$ \tilde\varepsilon'_{k2}=\tilde\varepsilon'_{k1} + 30,
\ \tilde\omega_{D}=43,\ \tilde\varepsilon_{F}=500.
$$
In this case superconductivity is possible near both points
$\varepsilon'_{k1}=0$ and $\varepsilon'_{k2}=0$.\\

\section{OVERLAP OF THREE ENERGY BANDS ON THE FERMI SURFACE}

We consider a crystal with strong electron-phonon and Coulomb interactions and
assume that repulsive interaction between electrons (holes) dominates.
Superconductivity could not occur in this case in a system with one energy
band. In multiband superconductors, where two or more energy bands overlap on
the Fermi surface, a system with repulsive interaction between electrons
(holes) becomes unstable to superconducting pairing. The cause of
superconductivity in this case is interband repulsion between electrons
(holes). We start with the Hamiltonian (2.1), where the first term corresponds
to the kinetic energy of the electrons of the $n$ bands and the second term
determines the interband repulsion between carriers ($ -V_{nm}>0$ for
$n\not = m$) with opposite momenta and spins. Because of the interband
interaction, the migration of a pair from one band to another and back always
leads to a net attractive interaction between the pairs and therefore even when
all the interaction constants are repulsive, interband interactions can lead to
a net attraction \cite{Moskalenko}-\cite{Konsin} (see Sec. 2). As noted above,
this is a qualitative difference between the multiband theory and the
one-band theory.  Therefore in this Section we consider the interband
interaction explicitly. In a more rigorous treatment it would be necessary to
take into account repulsive interaction between electrons inside each band
($-V_{nn} > 0$), which negatively affects superconductivity. Estimates of the
critical temperature $T_{c}$ in the two-band model \cite{Moskalenko},
\cite{Moskalenko_5} (Sec. 2) with $-V_{nm}>0$ show that the observed $T_{c}$ in
high-temperature superconductors can be obtained even when $\vert V_{nm}\vert <
\vert V_{nn} \vert$ by extending the interaction region to a large part of
momentum space and by introducing a cut-off energy in determining $T_{c}$.

Three energy bands overlap on the Fermi surface in the yttrium ceramics
$YBa_{2}Cu_{3}O_{7-\delta}$ for $\delta=0$ (Ref. 15). Therefore $n,m=1-3$ for
this compound. Using the method of Green's functions \cite{Abrikosov_2},
we obtain from (2.1) a system of equations for the three order parameters
$$
\hspace{5.2cm}
\Delta_{n}=\sum_{m,\vec k}V_{nm}\frac{th\left(\frac{\beta
E_{m}(\vec k)}{2}\right)}{E_{m}(\vec k)}\Delta_{m},\hspace{4.5cm}(6.1)
$$
$$
\hspace{5.4cm}
E_{m}=[(\varepsilon_{m} - \mu)^{2} + \Delta_{m}^{2}]^{1/2}.
\hspace{4.89cm}(6.2)
$$
Near the superconducting transition temperature we have
$$
\Delta_{1}=\sum_{k}V_{12}\frac{th\left[\frac{\beta(\varepsilon_2 -
\mu)}{2}\right]}{\varepsilon_{2} - \mu}\Delta_{2} +
\sum_{k}V_{13}\frac{th\left[\frac{\beta(\varepsilon_3 -
\mu)}{2}\right]}{\varepsilon_{3} - \mu}\Delta_{3}.
$$
$$
\Delta_{2}=\sum_{k}V_{21}\frac{th\left[\frac{\beta(\varepsilon_1-\mu)}{2}
\right]}{\varepsilon_1 - \mu}\Delta_{1} +
\sum_{k}V_{23}\frac{th\left[\frac{\beta(\varepsilon_3-\mu)}{2}\right]}
{\varepsilon_3 - \mu}\Delta_{3},
$$

$$
\hspace{3.0cm}
\Delta_{3}=\sum_{k}V_{31}\frac{th\left[\frac{\beta(\varepsilon_1-\mu)}{2}
\right]}{\varepsilon_1 - \mu}\Delta_{1} +
\sum_{k}V_{32}\frac{th\left[\frac{\beta(\varepsilon_2-\mu)}{2}\right]}
{\varepsilon_2 - \mu}\Delta_{3}.\hspace{3.0cm}(6.3)
$$
Examining the electronic superconductivity, we write the dispersion law for
the $n$th band in the form
$$
\hspace{4.8cm}
{\bf \varepsilon_n}=-E_{n} - \frac{\hbar}{2m_n}(k_{x}^{2} + k_{y}^{2}),\
m_{n}>0.\hspace{3.9cm}(6.4)
$$
This equation is an expansion of the energy
near the critical energy $E_{n}$ ($\nabla\varepsilon_{n}=0$) corresponding to a
topological transition \cite{Lifshits}. Such points always exist in an energy
band, but they normally do not affect the properties of metals, but show up in
the presence of impurities or pressure. According to band calculations,
\cite{Mattheis} - \cite{Krakauer} they also exist in high-temperature
superconductors. A change in the oxygen composition in an yttrium ceramics
changes the position of the Fermi level so that the relation $E_{n}=\mu$ is
satisfied, which favors the formation of cylindrical cavities in the Fermi
surface. The two-dimensional nature of the band spectrum is determined by the
$CuO_{2}$ plane.

In (6.3) we transform the series over $\bf k$ to integrals with respect to
energy with the help of the dispersion law (6.4). Let
$$
\hspace{6.0cm}
J(a,b)=\int_{a}^{b}\frac{th\left(\frac{\beta_c x}{2}\right)}{x}dx.
\hspace{5.3cm}(6.5)
$$
Then the system of equations (6.3) takes the form
\begin{eqnarray}
\Delta_{1}&=&N_{2}V_{12}J(\eta - E_{2},\eta - E_{c2})\Delta_{2}\nonumber\\
& &+N_{3}V_{13}J(\eta - E_{3},\eta - E_{c3})\Delta_{3},\nonumber
\end{eqnarray}
$$
\hspace{2.89cm}\Delta_{2}=N_{1}V_{21}J(\eta, \eta-E_{c1})\Delta_{1} +
V_{23}N_{3}J(\eta - E_{3},\eta - E_{c3})\Delta_{3},\hspace{2.0cm}(6.6)
$$
$$
\Delta_{3}=N_{1}V_{31}J(\eta,\eta - E_{c1})\Delta_{1} +
N_{2}V_{32}J(\eta - E_{2}, \eta - E_{c2})\Delta_{2}, $$ where $$ \eta=-\mu,\
N_{n}=\frac{k_{z}}{(2\pi)^{2}}\frac{m_{n}}{\hbar^{2}},
$$
$E_{cn}$ ($n=1,2,3$) are the cut-off energies in the corresponding bands. From
the condition for a nontrivial solution of the system (6.6) we obtain an
equation for the superconducting transition temperature $T_{c}=1/\beta_{c}$
\begin{eqnarray}
1&-&\lambda_{23}\lambda_{32}J(\eta - E_{2}, \eta -
E_{c2})J(\eta - E_{3}, \eta - E_{c3}) - \lambda_{12}\lambda_{21}\nonumber\\
& &\times J(\eta, \eta - E_{c1})J(\eta - E_{2}, \eta - E_{c2})\nonumber\\
& &-\lambda_{13}\lambda_{31}J(\eta, \eta - E_{c1})J(\eta - E_{3},
\eta - E_{c3}) \nonumber\\
& &- \lambda_{13}\lambda_{21}\lambda_{31}J(\eta,
\eta - E_{c1})J(\eta - E_{2},\eta - E_{c2})J(\eta - E_{3}, \eta - E_{c3})
\nonumber\\
& &-\lambda_{31}\lambda_{12}\lambda_{23}J(\eta, \eta -
E_{c1})J(\eta - E_{2}, \eta - E_{c2})\nonumber\\
& &\times J(\eta - E_{3}, \eta - E_{c3})=0;\nonumber\hspace{9.59cm}(6.7)
\end{eqnarray}
here $\lambda_{mn}=(N_{m}N_{n})^{1/2}V_{nm}$; $m,n=1,2,3$.

We simplify the problem by assuming that only the interband interactions
(1,2) and (2,3) are important, i. e. we put $\lambda_{13}=\lambda_{31}=0$. Then
\begin{eqnarray}
1 - \lambda_{12}\lambda_{21}J(\eta, \eta - E_{c1})J(\eta - E_{2}, \eta -
E_{c2})-\lambda_{23}\lambda_{32}J(\eta - E_{2}, \eta - E_{c2})& &\nonumber\\
\times J(\eta - E_{3}, \eta - E_{c3})=0.& &\nonumber\hspace{2.4cm}(6.8)
\end{eqnarray}
The condition $\lambda_{13}=\lambda_{31}=0$ can be satisfied when the
symmetries of the wave functions of bands 1 and 3 such that direct transitions
are forbidden. However, transitions  through the intermediate band are
possible and are taken into account. In this way, all three bands are active.
In a model with a large number of parameters, an assumption of this kind
simplifies the analysis without losing the effect of participation of all the
bands. In principle, the more complicated equation (6.7) could be considered
without difficulty.

It is of interest to study the dependence of $T_{c}$ on the parameter
$\eta=-\mu$ determining the position of the chemical potential. We assume
$a,b\gg T_{c}$ in the integral (6.5). Putting $E_{1}=0,\ E_{2}
< E_{3} < E_{c1} < E_{c2} < E_{c3}$, we obtain from (6.8)\\

1.  $\eta<E_{2}$:
\begin{eqnarray}
T_{c}&=&\frac{2\gamma}{\pi}\eta^{1/2}(E_{c1} - \eta)^{1/2}(E_{c3} -
\eta)^{\alpha}(E_{3}-\eta)^{-\alpha}\nonumber\\
& &\times\exp\left\{g - \frac{1}{2\chi_{12}\ln\left[\frac{(E_{c2} - \eta)}
{(E_{2} - \eta})\right]}\right\}.\nonumber\hspace{8.4cm}(6.9)
\end{eqnarray}
2. $\eta=E_{2}$:\\
\begin{eqnarray}
T_{c}&=&\frac{2\gamma}{\pi}(E_{c1} - E_{2})E_{2}^{1/4}(E_{c2} -
E_{2})^{-1/4}\nonumber\\
& &\times(E_{c3} - E_{2})^{\alpha/2}(E_{3} - E_{2})^{-\alpha/2}\nonumber\\
& &\times\exp\left\{-\frac{1}{2}\left[\ln^{2}\frac{E_{2}^{1/2}(E_{c3} -
E_{2})^{\alpha}}{(E_{c2}-E_{2})^{1/2}(E_{3} - E_{2})^{\alpha}} +
\frac{2}{\chi_{12}}\right]^{1/2} \right\}.\nonumber\hspace{4.17cm}(6.10)
\end{eqnarray}
3.  $E_{2} < \eta < E_{3}$:
\begin{eqnarray}
T_{c}&=&\frac{2\gamma}{\pi}\eta(E_{c1} - \eta)(\eta -
E_{2})(E_{c2} - \eta)]^{1/4}\nonumber\\ & &\times(E_{c2} -
\eta)^{\alpha/2}(E_{3} - \eta)^{-\alpha/2}\nonumber\\ &
&\times\exp\left\{-\frac{1}{2}\left[\ln^{2}\frac{(\eta - E_{2})(E_{c2} -
\eta)^{2}(E_{3} - \eta)^{2\alpha}}{(E_{c1} - \eta)^{2}(E_{c3} -
\eta)^{2\alpha}} + \frac{1}{\chi_2}\right]^{1/2}\right\}.\nonumber
\hspace{3.0cm}(6.11)
\end{eqnarray}
4.  $\eta=E_{3}:$
\begin{eqnarray}
T_{c}&=&\frac{2\gamma}{\pi}(E_{c1} - E_{3})^{1/4(1+\alpha)}\nonumber\\
&\times&(E_{3} - E_{2})^{1/4}(E_{c2} - E_{3})^{1/4}(E_{c3} -
E_{3})^{\alpha/2(1+\alpha)}E_{3}^{1/4(1+\alpha)}\nonumber\\
&\times&\exp\Bigg\{-\frac{1}{2}\Big[\frac{1}{4}\ln^2\frac{(E_3-E_2)
(E_{c2}-E_3)}{E_3^{1/(1+\alpha)}(E_{c1}-E_3)^{1/(1+\alpha)}
(E_{c3}-E_3)^{2\alpha/(1+\alpha)}}\nonumber\\
& &+\frac{1}{(\alpha+1)\chi_{12}}\Big]^{1/2}\Bigg\}.\nonumber
\hspace{10.2cm}(6.12)
\end{eqnarray}

5. $E_3 < \eta < E_{c1}$:
\begin{eqnarray}
T_c&=&\frac{2\gamma}{\pi}(E_{c2} - \eta)^{1/4}\nonumber\\
&\times&(\eta - E_2)^{1/4}\left[(E_{c1} - \eta)^{1/2}\eta^{1/2}(E_{c3} -
\eta)^{\alpha}(\eta - E_3)^{\alpha}\right]^{1/\alpha(1+2\alpha)}\nonumber\\
&\times&\exp\Bigg\{\frac{-1}{4(1 + 2\alpha)}\Big[\ln^2\frac{(E_{c2} -
\eta)^{1+2\alpha}(\eta - E_2)^{1+2\alpha}}{(E_{c1} -
\eta)(E_{c3}-\eta)^{2\alpha}\eta(\eta - E_3)^{2\alpha}}\nonumber\\
& &+\frac{4(1+2\alpha)}{\chi_{12}}\Big]^{1/2}\Bigg\},\nonumber\hspace{10.2cm}
(6.13)
\end{eqnarray}
where
$$
\chi_{12}=\lambda_{12}\lambda_{21},\ \chi_{23}=\lambda_{23}\lambda_{32},\
\alpha=\frac{\chi_{23}}{2\chi_{12}} .
$$

The dependence of $T_c$ on the filling of the energy bands $\eta=-\mu$
obtained from (6.9)-(6.13) is shown in Fig. 8.

For $YBa_2Cu_3O_{7-\delta}$ we can assume that when the Fermi level intersects
all three energy bands ($\delta=0$) the transition temperature reaches a
maximum $T_c\approx 90^{\circ} K$. As $\eta$ decreases ($\delta$
increases) $T_c$ drops and around $T=50^{\circ} K$ the dependence of $T_c$
on $\eta$ (or $\delta$) becomes weaker; the corresponding region in $\eta$ is
$1.2 \le \eta\le 1.9$ $eV$ for curve $2$. This dependence is
qualitatively consistent with the experimental curve of $T_c$ as a
function of $\delta$ (Ref. 47); compare Fig. 6. We note that the step-like
dependence of $T_c$ on $\delta$ was obtained in Ref. 57 in the two-band model
with repulsive interactions assuming that in a certain region $\mu$ does not
change with a charge in the concentration of oxygen.\\

\section{CONCLUSION}

We have summarized papers using the theory of superconductors with overlapping
bands to describe the properties of high-temperature superconductors. We have
also considered the effect of the Van Hove-Lifshits features of the density of
electron states. We have considered the formation of flat areas and
cylindrical cavities in the Fermi surface, which is typical of anisotropic
systems such as superconducting ceramics.

As noted above, in multiband systems superconductivity can occur even when all
the constants of the effective electron-electron interaction are repulsive.
Hence the analytical results in Secs. 2-5 can be used for $V_{nm}>0$ or
$V_{nm}<0$. The numerical estimates of the thermodynamic quantities in these
Sections are given for electron-phonon interaction with $V_{nm}>0$ (see
(2.1)).

Even in the weak coupling approximation the two-band approach predicts
a high value of $T_c$, a significant difference between $2\Delta_1/T_c$ and
$2\Delta_2/T_c$, a small value of the jump in the electronic heat capacity,
and so on (see Tables I and II). Note particularly the possibility of large
negative values of $d\ln T_c/d\ln V$, which have been observed experimentally
in lanthanum ceramics, and which represent an important test of the validity
of the theory.

It is interesting that the step-like dependence of $T_c$ on $\delta$ in
$YBa_2Cu_3O_{7-\delta}$ (Ref. 47) and also the two maxima in the pressure
dependence of $T_c$ in $Bi_2Sr_2CaCu_2O_8$ (Ref. 55) can be obtained assuming
the electron-phonon mechanism of superconductivity, if the overlap of the
energy bands and topological transitions are taken into account.

The three-band model considered in Sec. 6 gives still better agreement with
the experimental dependence of $T_c$ on $\delta$ in $YBa_2Cu_3O_{7-\delta}$.
In sec. 6 the basic cause of superconductivity was assumed to be interband
repulsive interactions between holes and in the equation determining $T_c$  the
integrals were cut off at energies of the order of the electron energy.

The results presented here show that the features of the band structure
(overlapping energy bands and topological transitions) play a crucial role in
determining the thermody-\\
namic and electromagnetic properties of high-temperature superconductors.

The model considered here takes into account the overlap of the energy bands
on the Fermi surface associated with the carriers in oxygen, which is
supported by NMR studies in yttrium ceramics (see Ref. 58, for example).
The NMR results show that the carriers in oxygen and in copper are two weakly
coupled subsystems which can be treated as independent in the first
approximation. We considered Coulomb repulsion in oxygen in the mean-field
approximation.  This approximation is standard and Coulomb interaction is
normally not taken into account, or where it is taken into account the
Hartree-Fock theory is used \cite{Hirch}, \cite{Yong}.

The important question of $p-d$ hybridization is not completely understood.
The experiments that stimulated the current research and which were compared
to theory for $YBa_2Cu_3O_{7-\delta}$ suggest that such hybridization is not
very significant.

The results obtained here are qualitative, since electron
correlations were not taken into account rigorously.

\end{document}